\documentclass[
    ,final            
  ]
  {aipproc}

\layoutstyle{6x9}

\begin{document}

\title{Strange Quark Contribution to the Nucleon Spin from Electroweak Elastic Scattering Data}


\author{S.F. Pate, J.P. Schaub\footnote{Present address:
The Evergreen State College, Olympia WA 98505, USA}, D.P. Trujillo}{
address={Physics Department, New Mexico State University, Las Cruces NM 88003, USA}
}

\begin{abstract}
The total contribution of strange quarks to the intrinsic spin of the nucleon can be determined from a 
measurement of the strange-quark contribution to the nucleon's elastic axial form factor. We have studied the strangeness 
contribution to the elastic vector and axial form factors of the nucleon, using elastic electroweak scattering data.  
Specifically, we combine elastic $\nu p$ and $\bar{\nu} p$ scattering cross section data from the Brookhaven E734 experiment 
with elastic $ep$ and quasi-elastic $ed$ and $e$-$^4$He scattering parity-violating asymmetry data from the SAMPLE, HAPPEx, G0 
and PVA4 experiments.  We have not only determined these form factors at individual values of momentum-transfer ($Q^2$), but 
also have fit the $Q^2$-dependence of these form factors using simple functional forms.  We present the results of these fits, 
along with some expectations of how our knowledge of these form factors can be improved with data from Fermilab experiments.
\end{abstract}

\maketitle



The techniques of inclusive and semi-inclusive polarized deep-inelastic scattering employed at CERN, SLAC, DESY,
and Jefferson Lab have provided a wealth
of information about the spin structure of the nucleon over the last 25 years.  The contributions of the
$u$ and $d$ quarks in the valence region have now been firmly established.  As well, data from collisions
of polarized protons at RHIC have done much to advance our knowledge of the limitations of the 
gluon spin contribution to the spin of the nucleon.  The strange quark contribution
to the spin of the nucleon, however, is still the subject of investigation, and the indications from deep-inelastic
scattering~\cite{Airapetian:2007mh, Airapetian:2004zf, Alekseev:2010ub} and 
from global fits~\cite{deFlorian:2009vb} are contradictory.  
It is of interest to examine the strange quark contribution to the nucleon spin
in a way that is independent of SU(3) symmetry and fragmentation functions.

The full strange quark contribution to the proton spin, $\Delta S$, 
can be directly determined by a measurement of the 
strange contribution to the proton elastic axial form factor, $G_A^s$,
in low energy electroweak elastic scattering.
$$\Delta S \equiv \Delta s + \Delta \bar{s} = G_A^s(Q^2=0)$$
By combining cross sections for $\nu p$ and $\bar{\nu}p$ elastic scattering with parity-violating asymmetries 
observed in $\vec{e}N$ elastic scattering, the strange quark contributions to the nucleon electromagnetic
and axial form factors $G_E^s$, $G_M^s$, and $G_A^s$ may be determined simultaneously~\cite{Pate:2008va,Pate:2003rk}.  

\begin{figure}[ht]
\begin{minipage}{18pc}
\includegraphics[trim = 0mm 50mm 0mm 0mm, clip, width=18pc]{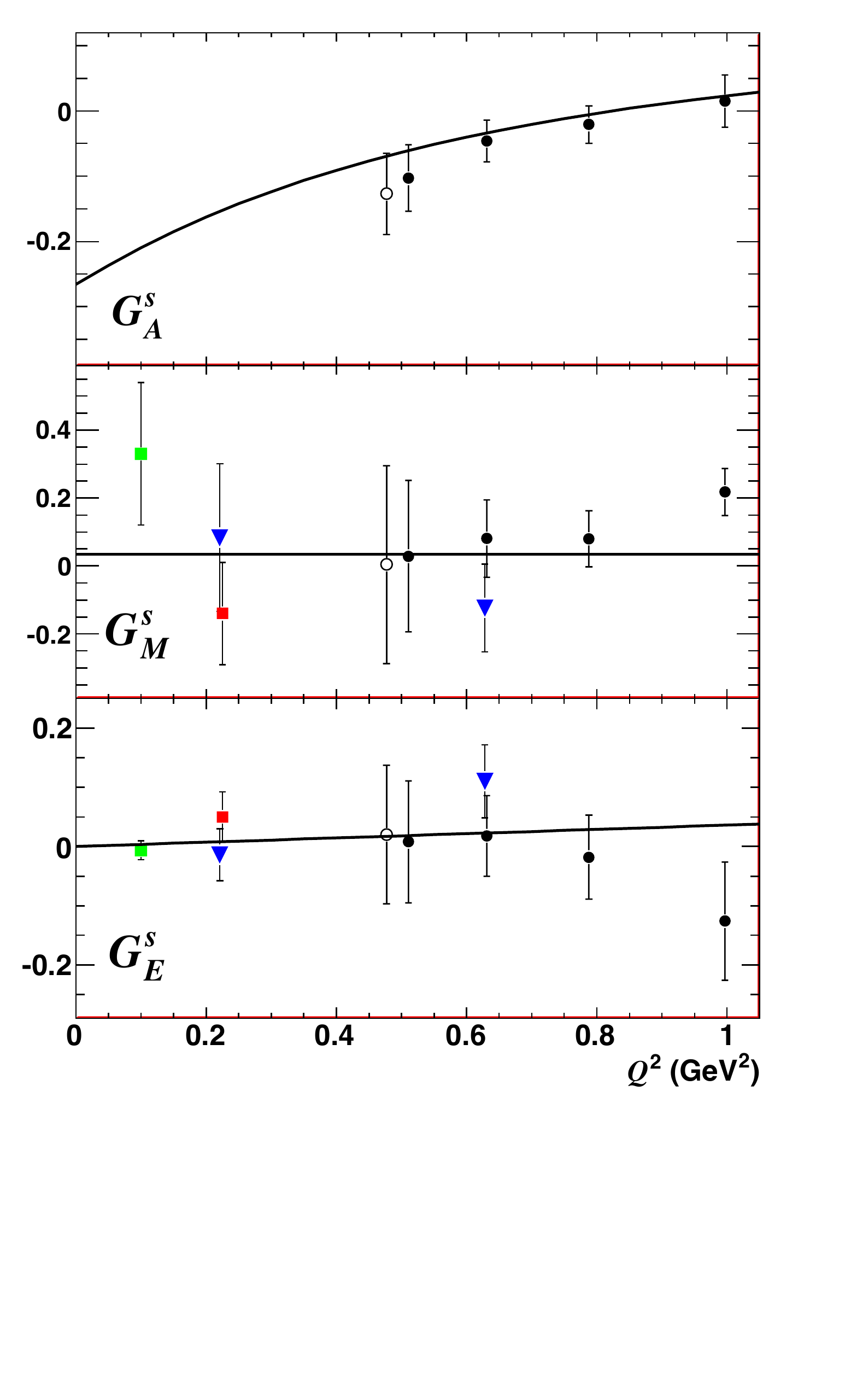}

\end{minipage}\hspace{2pc}%
\begin{minipage}{18pc}
\includegraphics[trim = 0mm 50mm 0mm 0mm, clip, width=18pc]{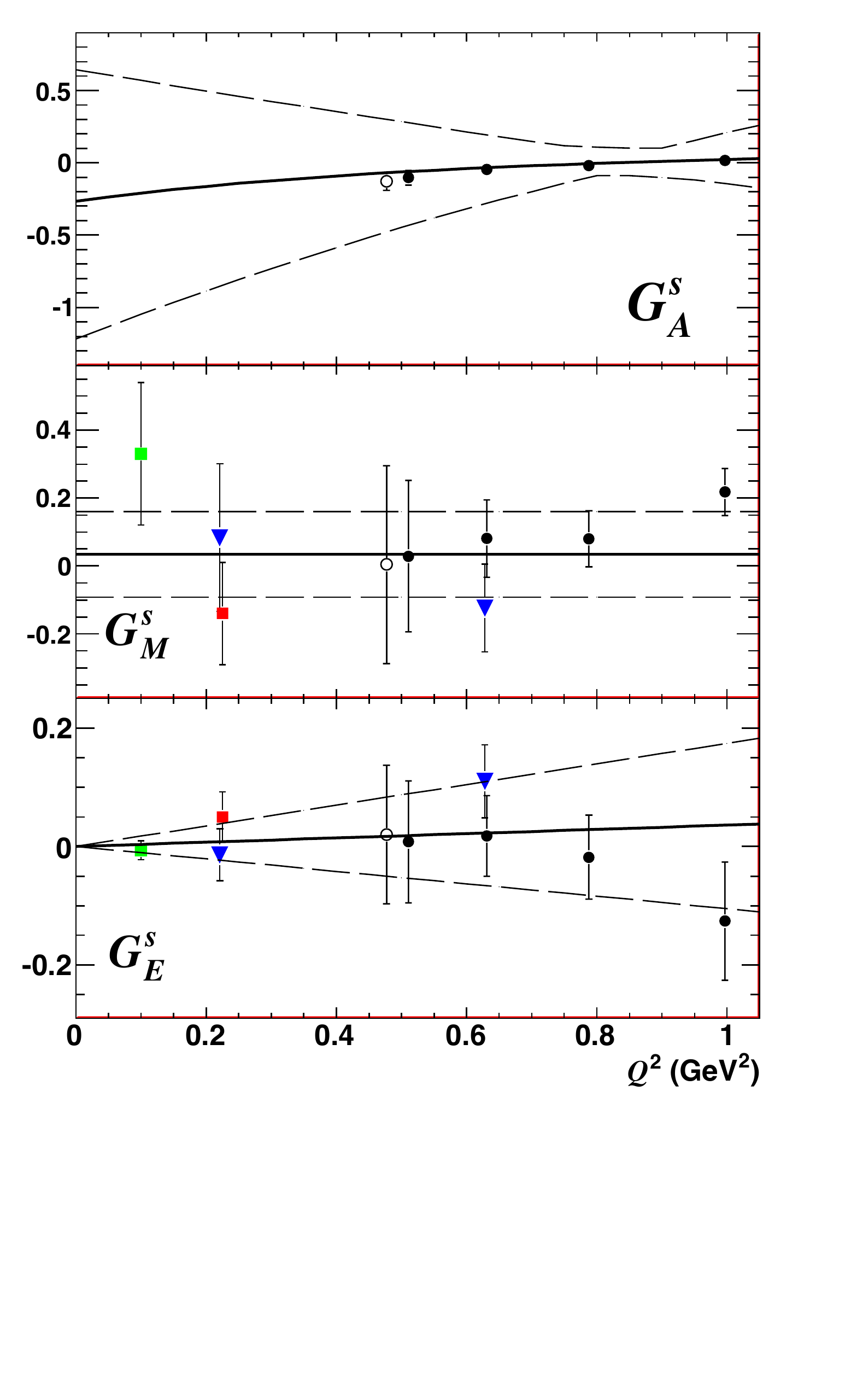}

\end{minipage} 
\caption{\label{fitfig} Results of the determination of $G_E^s$, $G_M^s$, and $G_A^s$ at
individual values of $Q^2$, and also from our global fit.  The separate determinations were done
by Liu et al.~\cite{Liu:2007yi} (green squares at 0.1 GeV$^2$), 
Androi\'{c} et al.~\cite{Androic:2009zu} (blue inverted triangles),
Baunack et al.~\cite{Baunack:2009gy} (red squares at 0.23 GeV$^2$), and
Pate et al.~\cite{Pate:2008va} (open and closed circles).  The preliminary results of the best global fit (see text)
are shown by the solid line; the 70\% confidence level limit curves for the fit are shown
as the dashed line in the right-hand panel.  The vertical scale for $G_A^s$ in the right-hand panel
has been adjusted to accommodate the limit curves of the fit.}
\end{figure}

A number of analyses have already determined $G_E^s$, $G_M^s$ at individual values of $Q^2$, and one of them
determines $G_A^s$ as well.  These results are reviewed in Figure~\ref{fitfig}.  
Each different analysis has taken some subset
of data covering a narrow range of $Q^2$, and uniquely determined the form factors based on standard model expressions
for the asymmetries and cross sections.  The strange quark contribution
to the {\bf vector} form factors is consistent with zero across the full range $0.1~{\rm GeV}^2<Q^2<1.0~{\rm GeV}^2$.  On
the other hand, there is some hint of a signal of a negative strangness contribution to the {\bf axial} form factor, 
but a lack of neutrino-scattering data
at low $Q^2$ prevents a definite conclusion about $\Delta S \equiv \Delta s + \Delta \bar{s} = G_A^s(Q^2=0)$ 
at this time.

We have performed a global fit of the available electroweak elastic scattering data, using the same general 
technique described in Ref.~\cite{Pate:2008va}, but now including all of the data from
the HAPPEx, G0, SAMPLE, PVA4, and BNL E734 experiments and assuming functional forms for the form factors.
A fit of this type can 
quantity the amount of information that can be extracted from the electroweak data, 
provide a mechanism for including new data that may become available,
and can be used as a tool to estimate the impact of any future experiments seeking to improve
our knowledge of the strangeness form factors.
  
We have fit the form factors $G_E^s$, $G_M^s$, and $G_A^s$ with this 
simple set of functional forms:
$$G_E^s = \rho_s\tau ~~~~~~~ G_M^s = \mu_s ~~~~~~~ G_A^s = \frac{\Delta S + S_A Q^2}{(1+Q^2/\Lambda_A^2)^2}$$
where $\tau = Q^2/4M_N^2$, $\rho_s \equiv (dG_E^s/d\tau)|_{\tau=0}$ is the strangeness radius, $\mu_s$ is
the strangeness magnetic moment, and $\Lambda_A$ defines how rapidly the $G_A^s$ function
depends on $Q^2$.
The linear term $S_A Q^2$ in the function for $G_A^s$ was needed to provide a good fit to the neutrino scattering data.
The best values for the five parameters (preliminary) are:
$$ \rho_s = 0.13 \pm 0.21 ~~~~~ \mu_s = 0.035 \pm 0.053$$
$$\Delta S = -0.27 \pm 0.41 ~~~~~ \Lambda_A = 1.3 \pm 1.9 ~~~~~ S_A = 0.32 \pm 0.48$$
The best fit is shown as the solid line in Figure~\ref{fitfig}; the 70\% confidence level uncertainty limits
are shown by the dashed line.

It should be stressed that the fit has {\bf not} been made to the values of the form factors displayed in 
Figure~\ref{fitfig}.  Instead, the fit has been made to the measured values of the cross sections and asymmetries
from the HAPPEx, G0, SAMPLE, PVA4, and BNL E734 experiments.  In Figure~\ref{fitfig}, we show 
that this global fit is consistent
with the results of previous local analyses of subsets of those data.

The preliminary results of the fit can be simply described.  The strangeness radius and magnetic moment 
are consistent with zero, and the uncertainties in these parameters are consistent with the uncertainties
in the separate determinations of $G_E^s$ and $G_M^s$ on display in Figure~\ref{fitfig}.  On the other hand,
$\Delta S$ is also consistent with 0 but the uncertainty is very large because there are no $\nu p$ or
$\bar{\nu} p$ elastic data at sufficiently low $Q^2$ to constrain it.  As a result the uncertainties in the
global fit to $G_A^s$ are very much larger than the uncertainties
in the separate determinations of $G_A^s$ in Figure~\ref{fitfig}.

It is clear that the single greatest need is for a much improved set of data on neutral current
$\nu p$ and $\bar{\nu} p$ elastic scattering down to the lowest $Q^2$ possible. 
MicroBooNE is an approved experiment at Fermilab to build a large liquid Argon Time Projection Chamber (LArTPC) 
to be exposed to the Booster neutrino beam and the NuMI beam at Fermilab. 
Liquid argon is very well-suited to observe low energy protons from low-$Q^2$ neutral-current and 
charged-current scattering; this experiment may be an ideal place to move towards a successful 
determination of $\Delta s$.

This work was funded by the US Department of Energy, Office of Science.



\bibliographystyle{aipproc}   

\bibliography{182_Pate}

\IfFileExists{\jobname.bbl}{}
 {\typeout{}
  \typeout{******************************************}
  \typeout{** Please run "bibtex \jobname" to optain}
  \typeout{** the bibliography and then re-run LaTeX}
  \typeout{** twice to fix the references!}
  \typeout{******************************************}
  \typeout{}
 }

\end{document}